\documentclass[]{emulateapj}

\usepackage{apjfonts,subfigure}

\def\LCDM{$\Lambda\mbox{CDM}$}

\def\GeV{{\rm\thinspace GeV}}

       \def\Mpc{{\rm\thinspace Mpc}}
\def\Msun{\hbox{$\rm\thinspace M_{\odot}$}}

\def\Msunpc2{{\Msun pc^{-2}}}

\shorttitle{Evidence for smooth accretion onto DM halos}
\shortauthors{Genel et al.}

\begin{document}

\title{The growth of dark matter halos: evidence for significant smooth accretion}

\author{Shy Genel\altaffilmark{1}, Nicolas Bouch{\'e}\altaffilmark{1,2}, Thorsten Naab\altaffilmark{3,4}, Amiel Sternberg\altaffilmark{5}, Reinhard Genzel\altaffilmark{1,6}}

\altaffiltext{1}{Max Planck Institut f\" ur extraterrestrische Physik, Giessenbachstrasse, 85748 Garching, Germany; shy@mpe.mpg.de; genzel@mpe.mpg.de}
\altaffiltext{2}{Department of Physics, University of California, Santa Barbara, CA 93106; nbouche@physics.ucsb.edu}
\altaffiltext{3}{Max Planck Institut f\" ur Astrophysik, Karl-Schwarzschild-Str. 1, 85741 Garching, Germany; naab@mpa-garching.mpg.de}
\altaffiltext{4}{Universit\"ats-Sternwarte M\"unchen, Scheinerstr.\ 1, D-81679 M\"unchen, Germany}
\altaffiltext{5}{School of Physics and Astronomy, Tel Aviv University, Tel Aviv 69978, Israel; amiel@wise.tau.ac.il}
\altaffiltext{6}{Department of Physics, Le Conte Hall, University of California, Berkeley, CA 94720}

\journalinfo{The Astrophysical Journal, accepted}

\begin{abstract}
We study the growth of dark matter halos in the concordance \LCDM{ }cosmology using several N-body simulations of large cosmological volumes. We construct merger trees from the Millennium and Millennium-II simulations, covering the ranges $10^{9}-10^{15}\Msun$ in halo mass and $1-10^{5}$ in merger mass ratio. Our algorithm takes special care of halo fragmentation and ensures that the mass contribution of each merger to halo growth is only counted once. This way the integrated merger rate converges and we can consistently determine the contribution of mergers of different mass ratios to halo growth. We find that all resolved mergers, up to mass ratios of $10^5:1$, contribute only $\approx60\%$ of the total halo mass growth, while major mergers are subdominant, e.g.~mergers with mass ratios smaller than $3:1$ ($10:1$) contribute only $\approx20\%$ ($\approx30\%$). This is verified with an analysis of two additional simulation boxes, where we follow all particles individually throughout cosmic time. Our results are also robust against using several halo definitions. Under the {\it assumption} that the power-law behaviour of the merger rate at large mass ratios can be extrapolated to arbitrarily large mass ratios, it is found that, independent of halo mass, $\approx40\%$ of the mass in halos comes from genuinely smooth accretion of dark matter that was never bound in smaller halos. We discuss possible implications of our findings for galaxy formation. One implication, assuming as is standard that the pristine intergalactic medium is heated and photoionized by UV photons, is that all halos accrete $>40\%$ of their baryons in smooth "cold" $T\gtrsim10^4K$ gas, rather than as warm, enriched or clumpy gas or as stars.

\end{abstract}


\section{Introduction}
\label{s:intro}
The way galaxies gain their mass affects almost every aspect of galaxy evolution. The distinction between, e.g., accretion of gas versus stars, spherical versus filamentary accretion, or clumpy versus smooth accretion will result in very different star-formation histories, colours, morphologies, angular momentum contents and sizes. Mergers are believed to play an important role in the evolution of galaxies, in particular of elliptical galaxies via the morphological transformation from disk-dominated galaxies to spheroids (e.g.~\citealp{ToomreA_77a,BarnesJ_92a,NaabT_03a,NaabT_06a,CoxTJ_06a,HopkinsP_07b,NaabT_09a,ConseliceC_03a,BellE_06b,LotzJ_08a}). Both theoretical and observational work have also emphasised the importance of smooth accretion of gas, in particular for the buildup of massive galaxies at high redshift and for the subsequent evolution of disk galaxies (e.g.~\citealp{WhiteS_91a,MuraliC_02a,KeresD_05a,OcvirkP_08a,DekelA_09b,GoerdtT_10a,DaddiE_07a,FoersterSchreiberN_09a,ConseliceC_09a,KauffmannG_10a}). The gas can be accreted in a spherically symmetric mode of cooling halo gas or in a filamentary mode directly from the cosmic web.

In the cold dark matter paradigm for structure formation \citep{BlumenthalG_84a}, galaxy formation is closely related to the formation of dark matter halos \citep{WhiteS_78a}, albeit in a complex way. Baryons follow the flow of the gravitationally-dominating dark matter and fall into dark matter halos, and galaxy mergers follow the mergers of their host dark matter halos. Understanding the process of halo mass assembly is an important step towards a better understanding of galaxy formation. In this paper we use high resolution dark matter simulations to study the relative importance of mergers versus smooth accretion for the buildup of halos and galaxies.

The non-linear nature of the evolution of gravitational instabilities into virialised structures makes N-body simulations the most reliable tool available for studying the mass buildup of dark matter halos. Massive dark matter particles in N-body simulations are believed to give a good representation of the coarse-grained phase-space structure of real dark matter. However, the interpretation of dark matter simulations is subject to some freedom and uncertainty in subsequent steps in making the connection to galaxies, such as the definition of a dark matter halo (e.g.~\citealp{WhiteM_01a,WhiteM_02a,CohnJ_07a,TinkerJ_08a}) and algorithms of merger tree construction.

One of the largest simulations used so far for studying the growth of dark matter halos is the Millennium Simulation (\citealp{SpringelV_05a}; hereafter MS). \citet{FakhouriO_07a} analysed the MS and found that the dark matter halo merger rate has a nearly universal form that can be separated into its dependencies on mass ratio, descendant mass, and redshift. They presented three algorithms for merger tree construction that result in merger rates differing by $\approx25\%$. In \citet{GenelS_09a} (hereafter G09) we introduced a novel merger tree construction algorithm we termed "splitting" (see also \citealp{FakhouriO_08b} and \citealp{MallerA_06a}) that incorporates the complicated process of halo fragmentation and re-merging such that a given pair of halos is never considered to merge more than once. This resulted in a different set of parameters for the \citet{FakhouriO_07a} global fitting formula for the merger rate extracted from our trees of the MS.

Thanks to ever increasing computation power, the dynamic range of N-body simulations increases too. As a result, several recent works have studied the growth rate of dark matter halos. When \citet{StewartK_08a} integrated the merger contribution and extrapolated to include {\it all} unresolved mergers, they found that only $\approx50-70\%$ of the final mass of halos was assembled by mergers. Rather dissimilar results were found by \citet{MadauP_08a}, who investigated the formation of one halo in a cosmological 'zoom-in' simulation, and by \citet{AnguloR_09a}, who investigated the growth of dark matter halos using high-resolution extended Press-Schechter (EPS; \citealp{PressW_74a,BondJ_91a,BowerR_91a}) trees. These works concluded that the mass accretion of dark matter halos is largely dominated by mergers. In this paper we perform a similar analysis to that of \citet{StewartK_08a} with a larger dynamic range using a combination of the MS and the higher resolution Millennium-II Simulation (\citealp{Boylan-KolchinM_09a}; hereafter MS2), and emphasise the importance of choosing an appropriate merger tree construction algorithm. We show that all resolved mergers contribute $\lesssim60\%$ to halo mass growth, and suggest that $\approx40\%$ of the accretion rate may be genuinely smooth. We also provide further support for these results by directly following the histories of dark matter particles in two smaller cosmological boxes, and make detailed comparisons with the previous results mentioned above.

This paper is organised as follows. In \S\ref{s:merger_trees_method} we review the Millennium simulations, structure identification and merger-tree construction. In \S\ref{s:merger_rate} we provide a fitting function that describes the halo merger rate, and in \S\ref{s:halo_growth} we discuss the contribution of mergers and smooth accretion to halo mass buildup. In \S\ref{s:comparison_EPS} we compare our results to the EPS model. In \S\ref{s:particles_method} we describe our direct analysis of dark matter particles histories in two additional simulations, and in \S\ref{s:particles_results} we present the results of that analysis. We discuss implications of our results to galaxy formation in \S\ref{s:galaxy_formation} and compare our results to previous work in \S\ref{s:previous_work}. We summarise in \S\ref{s:summary}.

\section{Analysis of merger trees}
\label{s:merger_trees}
\subsection{Method}
\label{s:merger_trees_method}
The MS is a cosmological N-body simulation that follows $2160^3$ dark matter particles, each of mass $8.6\times10^8h^{-1}\Msun$, in a periodic box of $500h^{-1}\Mpc$ on a side. The cosmology is \LCDM{ }with $\Omega_m=0.25$, $\Omega_{\Lambda}=0.75$, $\Omega_b=0.045$, $h=0.73$, $n=1$ and $\sigma_8=0.9$. The MS2 uses the same cosmology and follows the same number of particles, but in a box $5$ times smaller on a side. Thus, the MS2 particle mass is $125$ times smaller, i.e.~$6.885\times10^6h^{-1}\Msun$.

Structure identification in the simulations proceeds in two steps. First, the Friends-Of-Friends (FOF) algorithm (\citealp{DavisM_85a}; with a linking length parameter $b=0.2$) creates at every snapshot a catalogue of FOF groups that are considered to represent dark matter halos. Second, the algorithm SUBFIND \citep{SpringelV_01} identifies subhalos inside FOF groups by finding gravitationally self-bound collections of particles around maxima in the smoothed density field. The terminology is such that even smooth FOF groups with no identified substructures contain one {\it subhalo} (often referred to in the literature as the "main" or "background" subhalo) when they are self-bound. Therefore, FOF groups that contain zero subhalos are not gravitationally bound and are subsequently dropped from the analysis. Publicly available subhalo merger trees\footnote{Structure catalogues and merger trees were made public by the Virgo Consortium: http://www.mpa-garching.mpg.de/millennium.} were constructed by finding a single descendant for each subhalo, a procedure in which the FOF groups themselves played no role. However, if the halo merger rate is to be studied, then different types of merger trees need to be built, in which each node is a halo rather than a subhalo. Such a well-defined halo merger tree can be constructed if each FOF group is given exactly one descendant. In practice, however, choosing the correct descendant is not trivial because FOF groups not only merge, but may also fragment back into several groups.

In most fragmentation events, the subhalo (or group of subhalos) that left its original FOF group becomes a new distinct FOF group. Regardless of whether the two merge back together (as is usually the case) or not, a sequence of a merger followed by a fragmentation introduces artificial effects to the statistics of the merger rate and mass accretion. First, such a sequence of events is recorded as a merger even though the two progenitors end up as two distinct halos at the end of the merger-fragmentation sequence. Thus, the merger rate is overestimated. Second, a mass accretion rate of $M_{small}/\Delta t$ is attributed to the false merger ($M_{small}$ being the mass of the smaller companion and $\Delta t$ the time difference between snapshots), thus the contribution of mergers to halo growth is overestimated too. Third, if no "fragmentation rate" is quantified in parallel with the merger rate, all mass changes not associated with mergers are considered "smooth", and so a {\it negative} contribution of $-M_{small}/\Delta t$ is added to the smooth accretion component at the time of the fragmentation. Thus, the contribution of smooth accretion to halo growth is underestimated. Additionally, in trees where each halo is allowed to have one descendent at most (the standard case in the literature), the smaller product of the fragmentation has no progenitor, and so its main progenitor track is 'snipped' and all the information of its past formation history is in practice erased.

Therefore, an algorithm that builds fragmentation-free merger trees is needed to quantify correctly both the merger rate and the relative contributions of mergers and smooth accretion to halo mass growth. In G09 we presented such an algorithm and built new trees for the MS. Here we implement this algorithm on the MS2 as well. We construct the trees by splitting certain FOF groups, those that will suffer a fragmentation in the future, into several fragments. All the new fragments that our algorithm creates, as well as untouched FOF groups, are considered hereafter simply as "halos". A unique descendant is found for every halo, so that the merger tree is well defined. More details and motivation for our "splitting" algorithm and comparisons to other algorithms ("snipping", "stitching" and variants, as well as combinations, thereof) are presented in G09 and in \citet{FakhouriO_08b}. It is worth noting that already \citet{MallerA_06a} used a combination of "stitching" and "splitting" methods for their N-body/SPH simulation to obtain a fragmentation-free merger tree of galaxies. The halo merger trees we built (for the MS) are available at http://www.mpe.mpg.de/ir/MillenniumMergerTrees/.

\begin{figure*}[tbp]
\centering
\subfigure[]{
          \label{f:merger_rate_fits_1}
          \includegraphics[]{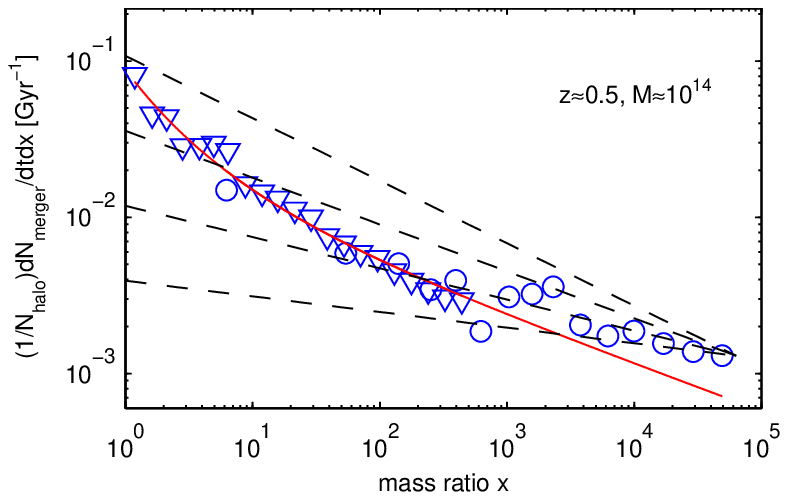}}
\subfigure[]{
          \label{f:merger_rate_fits_2}
          \includegraphics[]{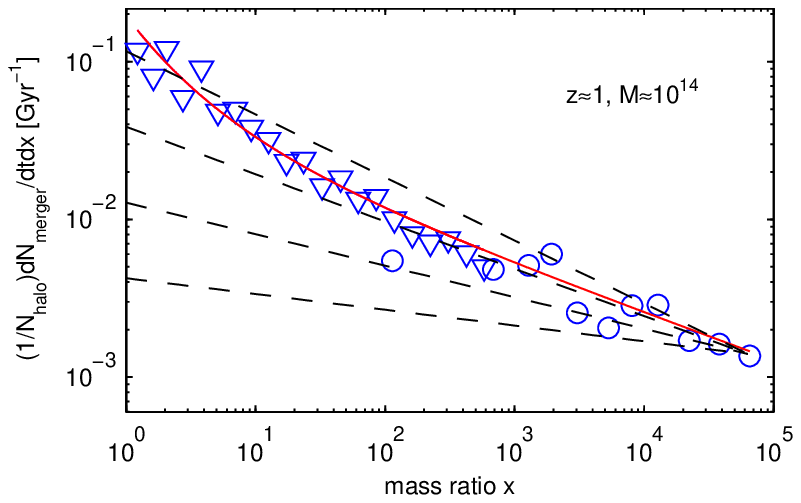}}
\subfigure[]{
          \label{f:merger_rate_fits_3}
          \includegraphics[]{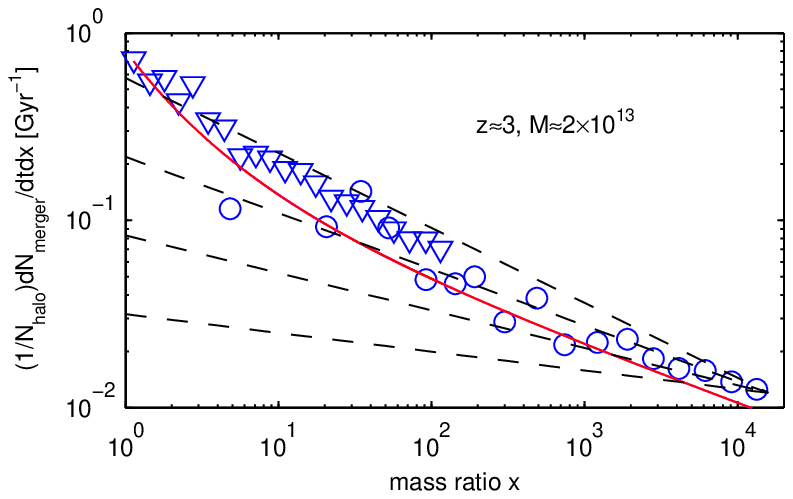}}
\subfigure[]{
          \label{f:merger_rate_fits_4}
          \includegraphics[]{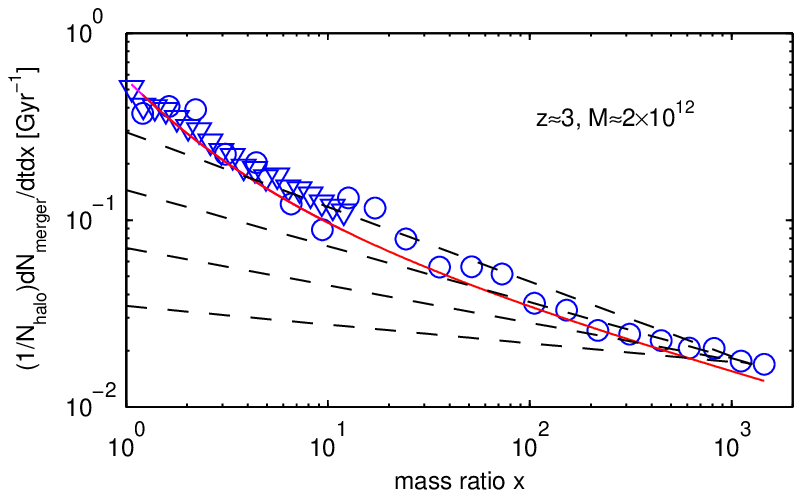}}
\caption{The merger rate per unit time from the MS ({\it triangles}) and the MS2 ({\it circles}). The different masses and redshifts are indicated in each panel. Our global fit (Equation \ref{e:rate_desc}), which has an asymptotic power-law index of $b=-0.3$ at high mass ratios, is shown by red solid lines. In each panel constant power-law indices of $\{-0.4,-0.3,-0.2,-0.1\}$ (from top to bottom) are shown by dashed black lines, and demonstrate that $b<-0.2$ is never a good fit to the data at $z\gtrsim0.5$. At redshifts $z\lesssim0.5$ our splitting algorithm cannot work properly due to the proximity to the end of the simulation: spurious mergers exist in the trees and the power-law index $b$ increases towards zero. Therefore results for low redshift are not shown here, and were not considered for the global fit.}
\vspace{0.5cm}
\label{f:merger_rate_fits}
\end{figure*}

We derive the merger rate per descendant halo per mass ratio $x$ per unit time $\omega\approx1.69/D(z)$, which is the natural time variable in the EPS model. Here, $D(z)$ is the linear growth rate of density fluctuations, and $\omega$ is estimated using the \citet{NeisteinE_08a} approximation. In our bookkeeping a merger between two halos of masses $M_1$ and $M_2\leq M_1$ is recorded as a merger at mass $M_1+M_2$ with ratio $x=M_1/M_2$. We define the halo mass as the mass of all particles gravitationally bound to it, i.e.~the sum of its subhalo masses (see G09 and \citealp{FakhouriO_09a}).

\subsection{Results from the merger trees}
\label{s:merger_trees_results}
\subsubsection{The merger rate}
\label{s:merger_rate}
We find that the merger rates in the MS and MS2 agree very well in the range of overlap ($M\approx10^{12}-10^{14}\Msun$). While the MS2, simulating a smaller volume, has worse statistics in that range, combining it with the MS provides a much larger dynamic range (see below). We fit the merger rate using the fitting formula introduced by \citet{FakhouriO_07a} (albeit with our mass ratio variable $x=1/\xi$). The new best-fitting parameters we find are only slightly different from those we found for the MS alone in G09. The fitting formula is
\begin{eqnarray}
\frac{1}{N_{\rm desc-halo}}\frac{dN_{\rm merger}}{d\omega dx}(x,z,M)=AM_{12}^\alpha x^b exp((\tilde{x}/x)^\gamma)
\label{e:rate_desc}
\end{eqnarray}
where $M_{12}=M/10^{12}\Msun$. Our best-fitting parameters for the combination of the two simulations are: $A=0.065$, $\alpha=0.15$, $b=-0.3$, $\tilde{x}=2.5$ and $\gamma=0.5$ \footnote{With the mass ratio definition used by \citet{FakhouriO_07a} our parameters correspond, in their notation, to: $A=0.065$, $\alpha=0.15$, $\beta=-b-2=-1.7$, $\tilde{\xi}=1/\tilde{x}=0.4$, $\gamma=0.5$, $\eta=1$ and $\tilde{M}=10^{12}\Msun$.}.

Figure \ref{f:merger_rate_fits} shows a few examples of the merger rate for different halo masses and redshifts and the corresponding fits. We find our fit to be appropriate for the whole mass range probed by the simulations at redshifts $0.5\lesssim z\lesssim5$. At $z\gtrsim5$ the redshift and mass dependencies become stronger, and we do not attempt to fit that regime. At $z\lesssim0.5$ the fit fails due to the proximity to the end of the simulations. That is, the subhalo disruption time at $z\lesssim0.5$ becomes comparable to the lookback time corresponding to that redshift, thus we cannot identify all the artificially connected halos that would have fragmented had the simulations run past $z=0$. Figure \ref{f:merger_rate_fits} shows that the power-law index $b$ that describes the merger rate at large mass ratios is robustly constrained to be $b\lesssim-0.2$. The important consequences of this finding will be discussed in \S\ref{s:halo_growth}.

The only regime where there is a significant difference between results from the MS and the MS2 is where halos of $<100$ particles in the MS are involved. In G09 we used a lower threshold of $40$ particles, and found an upturn of the merger rate when the less massive halo had between $40$ and $\approx100$ particles. Mergers involving the same halo masses in the MS2 are resolved with many more particles, and no such upturn appears there. This is qualitatively understandable given the finding of \citet{WarrenM_06a} that FOF groups with low particle numbers are overestimated in mass. Hence, combining the two simulations shows that the lower threshold of $40$ particles we used in G09 is too low, but that mergers involving two halos of $>100$ particles are well resolved. Therefore, we have excluded mergers with halos that consist of less than $100$ particles from our global fit. This places a limit for the halo mass of $M\approx1.2\times10^{11}\Msun$ for the MS and $M\approx9.4\times10^{8}\Msun$ for the MS2. Given that there are enough statistics for halos of $M\approx10^{15}\Msun$ in the MS and of $M\approx10^{14}\Msun$ in the MS2, the largest merger mass ratios we can reliably probe are $\approx10^{4}$ with the MS and $\approx10^{5}$ with the MS2.

\subsubsection{Halo growth modes}
\label{s:halo_growth}
\begin{figure}[tbp]
\centering
\includegraphics[]{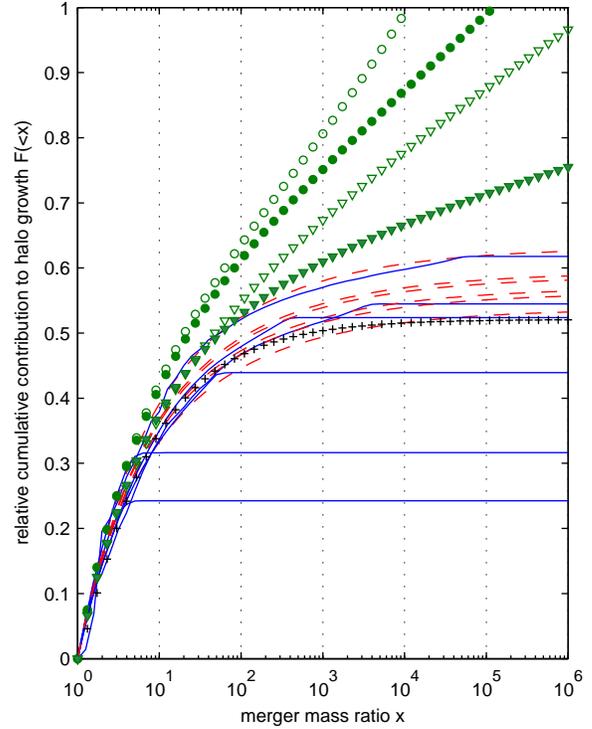}
\caption{The relative contribution of mergers to halo mass growth $F(<x)$ as a cumulative function of mass ratio $x$. The solid blue curves describe $F(<x)$ for $M\approx10^{11},10^{13},10^{14}$ from the MS (going up to mass ratios of $4,400,4000$, respectively) and $M\approx10^{9},10^{10},10^{13}$ from the MS2 (mass ratios of $5,50,50000$, respectively). The total contribution of mergers is at most $\approx60\%$ of the total growth rate of halos. The dashed red curves show the integral of the merger rate (equation (\ref{e:rate_int})) using the "splitting" trees and normalising to the total actual growth. The green symbols show the same quantity, for $M\approx10^{12}$ in the MS, for other merger tree construction algorithms: "snipping" from \citet{FakhouriO_07a} ({\it circles}), "stitching-3" ({\it filled circles}) and "splitting-3" from \citet{FakhouriO_09a} ({\it triangles}), "splitting-3" from \citet{FakhouriO_10a} ({\it filled triangles}), and the combined method of \citet{StewartK_08b} ({\it pluses}).}
\vspace{0.5cm}
\label{f:growth}
\end{figure}

In the following we investigate the relation between the total mass growth of halos, the relative mass growth due to mergers and the halo merger rate. In Figure \ref{f:growth} the solid blue curves show $F(<x)$, the fractional cumulative contribution of mergers to the total instantaneous growth rate of halos. Those contributions are summed up directly from all mergers in our trees. There are $6$ such curves, $3$ from the MS and $3$ from the MS2, for masses spanning the range $10^{9}\Msun$ to $10^{14}\Msun$, and averaged over the redshift range $1<z<3$ (we find only a weak redshift dependence of $F(<x)$). Each solid curve breaks at some $x$ and becomes horizontal - this is the mass ratio above which mergers cannot be resolved anymore, depending on the mass bin and simulation used\footnote{Here, as opposed to the case of the merger rate, we do show the contribution of mergers with {\it all} halos, i.e.~down to the resolution limit of $20$ particles. While halos with less than $100$ particles show an upturn in the merger rate, their influence on the mass contribution is very small, and so we include them in Figure \ref{f:growth} in order to show the full contribution of all mergers in the simulation. The very small upturn at the higher mass ratios in some of the blue curves in Figure \ref{f:growth} are evidence for this upturn.}.

The solid blue curves in Figure \ref{f:growth} form together a common envelope in the range where mergers are resolved. This envelope shows that $\approx20\%$ of the total growth rate comes from major mergers ($1<x<3$), mergers with $1<x<10$ contribute $\approx30\%$ of the mass growth, all mergers with $1<x<100$ contribute only $\approx45\%$, and the total relative mass contribution of mergers even in the best resolved case, up to a mass ratio of $10^{5}$, is no more than $60\%$.

Figure \ref{f:growth} also shows the integration of the merger rate obtained by implementing other algorithms for merger tree construction. The black pluses are for the merger rate fitting formula provided by \citet{StewartK_08b}. Their method results in a converging merger contribution that agrees well with ours\footnote{The merger rate quantified by \citet{StewartK_08b} has some notable differences to ours in its mass ratio and redshift dependencies, but they become much less significant when the fractional cumulative contribution is considered, as in Figure \ref{f:growth}.}, because they pay attention not to double-count mergers. To do that, they use a combination of the "stitching-$\infty$" and splitting algorithms (see G09 and \citealp{FakhouriO_08b} for a detailed comparison of the different algorithms). The green symbols are for methods where some fragmentations remain in the trees. These are "snipping" (which is equivalent to not treating fragmentations at all; {\it circles}) from \citet{FakhouriO_07a}, "stitching-3" ({\it filled circles}) and "splitting-3" from \citet{FakhouriO_09a} ({\it triangles}), as well as "splitting-3" from \citet{FakhouriO_10a} ({\it filled triangles}). The former two have an asymptotic power-law with $b>0$, which means that the total merger mass contribution diverges as $x$ increases. The latter two converge, but still show a very different shape from what is obtained from our trees. This does not mean that those methods do not conserve mass. Rather, as mergers with increasing $x$ are resolved, their artificial contribution due to fragmentation, as described in \S\ref{s:merger_trees_method}, increases, while the compensation comes in the form of negative contributions from smooth accretion. When mergers with high enough $x$ are resolved, those methods are expected to give negative smooth accretion rates.

The conclusion is that in merger trees that are built so that some fragmentations remain, halo mass assembly must be described by three components: the merger rate, the smooth accretion rate and the fragmentation rate. Otherwise, the interpretation of "anything but mergers" as "smooth" is false. In \S\ref{s:particles} we use a particle-based analysis that is independent of the merger tree construction algorithm to show that the contribution of mergers is consistent with our fragmentation-free "splitting" trees.

The finding that at most\footnote{Note that $F(<x)$ in Figure \ref{f:growth} is averaged over different halos. The {\it instantaneous} value for individual halos may be very different.} $60\%$ of the growth rate of halos is achieved via mergers with $1<x<10^{5}$ is already remarkable. But what if we had a simulation with an even larger dynamic range? We estimate this by using an extrapolation of the merger rate. The mass growth due to mergers with $x_1<x<x_2$ is
\begin{eqnarray}
&\int_{x_1}^{x_2}\frac{1}{N_{\rm desc-halo}}\frac{dN_{\rm merger}}{d\omega dx}(x,z,M)M_{small}dx,
\label{e:rate_int}
\end{eqnarray}
where $M$ is the descendant mass and $M_{small}$ is the mass of the less massive progenitor of each merger. Evaluating the integral of the merger rate per descendant halo requires specifying $M_{mp}/M$, where $M_{mp}$ is the main progenitor mass, since the merger mass ratio is defined such that $M_{small}=M_{mp}/x$. We approximate $M_{mp}/M$ by $\frac{x}{1+x}\langle M_{mp}/M\rangle$, where $\langle M_{mp}/M\rangle$ is the mean $M_{mp}/M$ computed separately for each $M$ in each of the simulations and averaged over redshift. The dashed red curves in Figure \ref{f:growth} show equation (\ref{e:rate_int}) evaluated between $x_1=1$ and $x_2=x$ and normalised to the total actual growth. There is a very good agreement with the directly extracted fractions. This integration demonstrates that $F(<x)$ converges at $x\gg1$ to $\approx60\%$. The convergence can be easily understood, since the merger rate behaves as a power-law with $b=-0.3$ at $x\gg1$ (see equation (\ref{e:rate_desc})). Naturally, we cannot be certain that an extrapolation is valid. Yet, for $F(<x)$ to converge to $1$, more minor mergers are needed below the resolution limit, such that the asymptotic power-law index would have to be $b\approx-0.01$ at $x\gtrsim10^5$. Such an index, as we demonstrate in Figure \ref{f:merger_rate_fits}, is excluded by the data at the currently available resolution ($x\lesssim10^5$), thus for the fractional mergers contribution to converge to $100\%$, the power-law index of the merger rate must {\it change} below our resolution limit.

\subsubsection{Comparison to the EPS model}
\label{s:comparison_EPS}
\begin{figure}[tbp]
\centering
\includegraphics[]{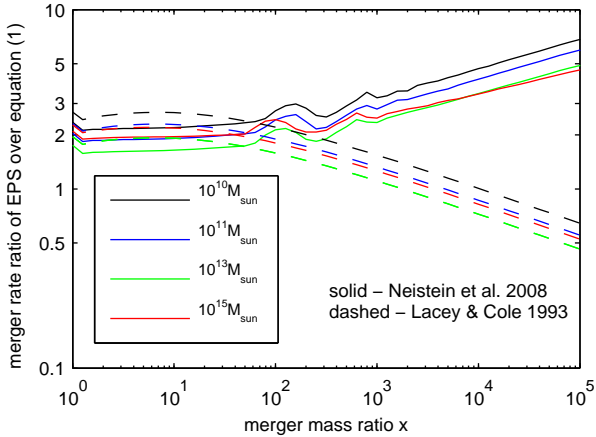}
\caption{The ratio of the merger rate predicted from EPS to our global fit. This ratio is independent of redshift. The EPS merger rate is higher, which roughly compensates for the lack of smooth accretion in the EPS model to give a similar total accretion rate.}
\vspace{0.5cm}
\label{f:EPS_ratio}
\end{figure}

Figure \ref{f:EPS_ratio} shows the ratio of the merger rate predicted from the EPS model by \citet{LaceyC_93a} (dashed) and \citet{NeisteinE_08a} (solid) to our global fit equation (1) for different masses (and independently of redshift). We identify two regions: at $x\lesssim100$ the ratio is almost independent of $x$ and ranges from $\approx1.6$ to $\approx2.3$ for different masses, while at $x\gtrsim100$ the ratio is a power-law. This is because our fit has an asymptotic power-law index $b=-0.3$, while the \citet{NeisteinE_08a} merger rate has a shallower index of $-0.16\pm0.01$ and that of \citet{LaceyC_93a} a steeper index of $-0.5$.

Since most of the merger-contributed accretion rate comes from $x<100$ (Figure \ref{f:growth}), and EPS has a $\approx100\%$ higher merger rate in that regime, the total accretion rate of EPS from $x<100$ equals almost the actual {\it total} accretion rate measured in the simulations. Since in the EPS model all the growth comes from mergers by construction (see however \citet{AnguloR_09a}), it seems that the EPS prediction differs from the simulation results in two ways that roughly cancel each other: smooth accretion is lacking, but this is compensated by a boosted merger rate. The higher merger rate found by \citet{NeisteinE_08a} in the $x>100$ regime boosts their self-consistent EPS total accretion rate further, so that it overestimates the total accretion rate in the simulations by $\approx35\%$, with weak dependencies on mass and redshift.

\section{Analysis of particle histories}
\label{s:particles}
\subsection{Method}
\label{s:particles_method}
As much as arguments exist in favour of one or the other algorithm for merger tree construction, some freedom is still left due to the complexity of the hierarchical buildup of cosmic structures. As the results we presented in \S\ref{s:merger_trees} (specifically Figure \ref{f:growth}) are algorithm-dependent, it is beneficial to perform an analysis that is independent of such algorithms. Comparing the results of such an analysis to the results from various merger trees can also serve as a tool for distinguishing between the algorithms. In this Section we present an analysis of particle histories that circumvents many of the details involved in building merger trees and just relies on the identification of structure and identification of a 'main progenitor trunk' for each halo. This direct particle analysis allows us to get a better handle on the nature of the smooth component.

We perform this analysis on two cosmological N-body simulations. One is the milli-Millennium Simulation that uses the same cosmology and has the same resolution as the MS but includes a factor of $512$ less particles in a box of $62.5h^{-1}\Mpc$ on a side. The second is the USM Simulation (first presented in \citealp{MosterB_09a}) that uses somewhat different cosmological parameters that are in better agreement with current observations ($\Omega_m=0.26$, $\Omega_{\Lambda}=0.74$, $\Omega_b=0.044$, $h=0.72$, $n=0.95$ and $\sigma_8=0.77$) and follows particles of mass $2\times10^8h^{-1}\Msun$ (i.e.~$4.3$ times smaller than in the MS) in a box $72h^{-1}\Mpc$ on a side. 

We distinguish between three modes of accretion: 'merger', 'smooth' and 'stripped'. In broad terms, we assign any particle accreted as part of a merger event as 'merger accretion', while 'smooth accretion' is the accretion of particles that never belonged to a bound structure earlier than the accretion event and 'stripped accretion' is the accretion of particles that do not arrive as part of a halo at the time of accretion but were part of an identified halo at some earlier time. More precisely, we follow each particle $p$ that belongs to any halo $h$ at any snapshot $s_0$ to the first snapshot $s_{acc}$ at which it belonged to the main progenitor trunk of halo $h$. The halo on the main progenitor trunk of halo $h$ at snapshot $s_{acc}$ is termed $h_{acc}$. Note that, as we discuss below, particles may 'cycle' in and out of their halos, i.e.~particle $p$ does not necessarily belong to the main trunk of halo $h$ at all snapshots $s_0>s>s_{acc}$, but we are interested in the 'accretion mode' of $p$ at the very first time it belonged to the main progenitor trunk of $h$. We then look for particle $p$ in snapshot $s_{acc}-1$, and tag it according to the following criteria. If $p$ at $s_{acc}-1$ belongs to a progenitor halo of $h_{acc}$, it is tagged as 'merger accretion'. If it belongs to a halo that is not a progenitor of $h_{acc}$, it is tagged as 'stripped accretion'. If $p$ belongs to no halo at $s_{acc}-1$, we follow it back through every snapshot to the initial conditions. If we find some earlier snapshot $s<s_{acc}-1$ where $p$ belonged to a halo, we also tag it as 'stripped accretion', otherwise it is tagged 'smooth accretion'.

Further on we tag some particles as 'leaving' or 'joining' their halo. 'Leaving' particles are those that are not part of the halo's direct descendant (at snapshot $s=s_0+1$). 'Joining' particles are those that did not belong to the halo's direct main progenitor (at $s=s_0-1$). The net growth rate of each halo is then given by $N_{join}-N_{leave}$. Since we also know the original accretion mode of each particle onto the halo, we can quantify the contribution of each mode to the total net growth rate.

It is important to discuss to what extent the results from this analysis are expected to be independent of the merger tree on which the analysis is based. Let us examine the consequences of a fragmentation in the merger tree. The more massive halo of the pair, the one for which the main progenitor trunk remains intact, will be insensitive to whether the fragmentation is cured by splitting the spuriously-connected halo or not. This is because a fragmentation event of a subhalo (that previously arrived in a merger) will appear as 'leaving' particles tagged as 'merger mode' and therefore on average will cancel out the same particles when they were 'joining'. On the other hand, implications do exist for the other halo of the pair, i.e.~the smaller 'fragment', if the fragmentation is left in the tree. The small fragment has no progenitor and so its main progenitor trunk is cut and therefore the memory of its particles as for their true origin is erased. After the fragmentation they will all be classified as 'stripped', because they belong to a halo that is not the fragment's progenitor just prior to its 'appearance' as an independent halo. Therefore, we expect an artificial overestimate of 'stripped' particles if fragmentations are left in the tree, at the expense of both other accretion modes. We show and discuss this effect further in Section \ref{s:particles_results}.

\subsection{Results from the analysis of particles}
\label{s:particles_results}
\begin{figure}[tbp]
\centering
\includegraphics[]{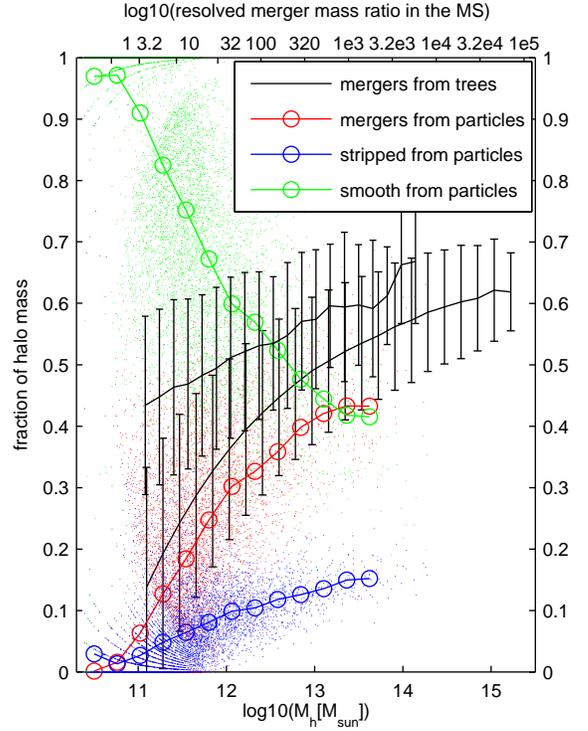}
\caption{The fraction of the particles in $z=0$ halos originating from the three accretion modes: 'merger' ({\it red}), 'stripped' ({\it blue}) and 'smooth' ({\it green}), versus halo mass. The dots are for different halos and demonstrate the distribution around the mean values, which are shown by curves with open circles. Overplotted are corresponding curves from the analysis of the merger trees of the MS ({\it lower black curve}) and MS2 ({\it upper black curve}), which indicate the total mass accreted via mergers onto the main progenitor trunk normalised to the $z=0$ halo mass. The error bars include $68\%$ of the halos around the mean (see \citealp{McBrideJ_09a} for a more detailed description of the distribution around the mean for major mergers). The MS curve from the merger trees is similar, yet slightly higher, than the 'merger' component from the particle analysis. This is expected since some particles that arrive via mergers leave (smoothly) and never come back, which is information that is not included in the merger trees alone. The MS2 curve is significantly higher where MS halos are barely resolved and MS2 halos are well resolved, but the difference between the two curves decreases towards higher halo masses, as expected. Note that the mass content of halos is only slightly affected by the proximity to the end of the simulation and the inability to split some subhalos out. This is because the mass ratio regime that is mostly affected by this effect is subdominant in mass contribution.}
\vspace{0.5cm}
\label{f:halo_content}
\end{figure}

Figure \ref{f:halo_content} shows the fractional mass content of $z=0$ halos in the milli-Millennium simulation separated into the three modes by which each of their particles was accreted onto them, as indicated in the legend. The dots are for different halos and demonstrate the distribution around the mean values, which are shown by curves with open circles. We see ({\it green}) that almost all halos are made mostly of particles that never belonged earlier to other halos. Plotted on top ({\it bottom black}) are results from the analysis of the merger trees, which are obtained by summing up all the mass accreted via mergers along a halo's main progenitor trunk. An important point to take from Figure \ref{f:halo_content} is the agreement between the two analysis methods, shown by the red and bottom black curves. The small difference between the two is expected because some particles that arrive via mergers later leave the halo and do not contribute to its mass at $z=0$ (see also below).

The trend seen in Figure \ref{f:halo_content} is clearly an effect of resolution: halos closer to the resolution limit have gained most of their mass by unresolved accretion, which cannot be distinguished between the different components and this mass is identified as 'smooth'. Nevertheless, the beginning of a saturation of the mass fraction arriving from mergers is apparent as halos are better resolved. We suggest that the 'true' value for all halos is the saturated value that is seen for high-mass halos. Additional evidence for the trend being an effect of resolution is the fact that for the same halo mass, the MS2 halos, which are better resolved, show a weaker trend ({\it top black}) than the results from the MS ({\it bottom black}). We do not have enough statistics to quantitatively constrain the saturation of this quantity due to low halo numbers at high masses, but these results are consistent with our results on the accretion rate (see below), where the saturation is statistically robust (\S\ref{s:halo_growth}).

In Figure \ref{f:particle_net_accretion} we show the main result of the particle analysis, the fraction of the net instantaneous accretion that is associated with the three different modes, as a function of halo mass and redshift. We observe no dependence on redshift, even down to $z=0$, despite the inflated merger rate that is due to the proximity to the end of the simulation. This means that even if some of the increase of the minor merger rate at $z\lesssim0.5$ is {\it real}, it is not high enough to make a significant contribution to the total mass accretion (see also Figure \ref{f:fraction_below_UV_threshold} below). The mass dependence is again a resolution effect. As halos are better resolved, the mergers they undergo are better resolved and so the contribution of the 'merger' component increases. Although we again cannot make a robust fit and extrapolation from these results, we find it reassuring that the particle analysis gives a consistent result with that of the "splitting" merger trees ({\it dashed black curves}, which cover the range of the dashed curves in Figure \ref{f:growth}).

It is worth noting that the curves in Figure \ref{f:halo_content} are shifted from those in Figure \ref{f:particle_net_accretion} by $\approx0.5-1 dex$ toward higher masses. This is expected, as halos accrete most of their final mass $M_{z=0}$, by definition, when $M_z\gtrsim0.1M_{z=0}$. This means that the fractions of the different accretion modes in the final halo mass $M_{z=0}$ (Figure \ref{f:halo_content}) correspond to their fractions in the accretion at $M\gtrsim0.1M_{z=0}$ (Figure \ref{f:particle_net_accretion}).

\subsection{The cycle of particles in and out of halos}
\label{s:particle_cycle}
\begin{figure}[tbp]
\centering
\includegraphics[]{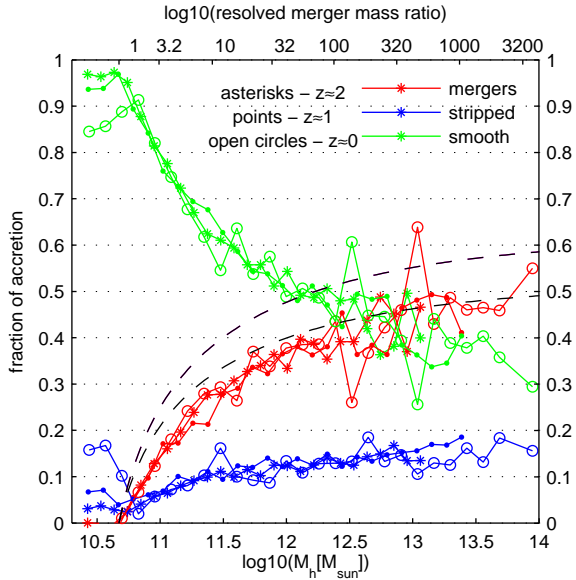}
\caption{The fraction of net accretion (i.e.~'joining' particles minus 'leaving' particles), as a function of halo mass, that belongs to each of the three modes 'merger' ({\it red}), 'stripped' ({\it blue}) and 'smooth' ({\it green}). The results are shown for three different redshifts and are independent of it. As higher mass ratios are resolved, more of the accretion is in the 'merger' and 'stripped' modes, but the merger mode contribution is consistent with the saturation inferred from the merger trees (shown by the dashed black curves that cover the range of the dashed curves in Figure \ref{f:growth}). This indicates that the contribution of mergers may not increase further.}
\vspace{0.5cm}
\label{f:particle_net_accretion}
\end{figure}

From Figures \ref{f:halo_content} and \ref{f:particle_net_accretion} we can also learn that the fraction of the accretion in the 'stripped' mode is subdominant to that in the 'smooth' mode. That is, what we could only interpret as 'non-mergers' from the analysis of the merger trees, can now be shown to consist of particles that never belonged to another halo prior to their accretion. In fact, the 'stripped' mass is consistently $\approx1/3$ of the mass in the 'merger' mode. Conservatively, the 'stripped' component is our uncertainty, because our analysis does not indicate how long it has been stripped and what part of it comes from the vicinity of approaching subhalos. Indeed, some of these 'stripped' particles arrive as mergers into low-mass halos and after being stripped from them, are re-accreted into more massive halos and tagged there as 'stripped'. Thus, some of the mass appearing in the merger trees as 'merger mode' is transferred into 'stripped mode' of more massive halos in the particle analysis. This is the reason the red curves are somewhat lower than the black dashed ones.

\begin{figure}[tbp]
\centering
\includegraphics[]{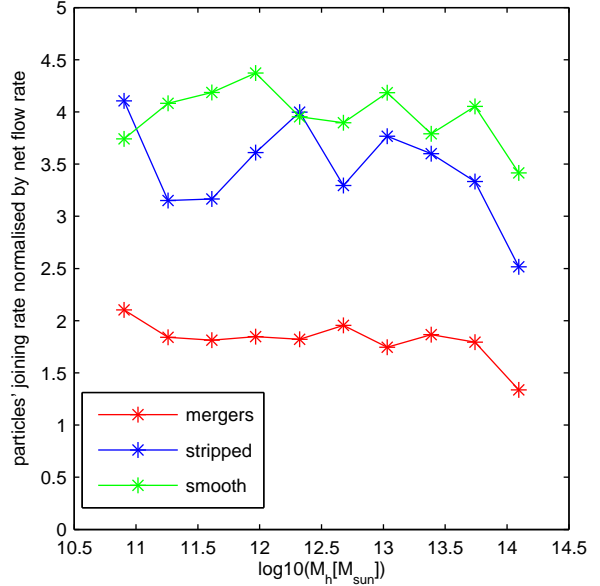}
\caption{The $z=0$ cycle of 'joining' and 'leaving' particles, shown by the ratio of the gross inflow rate ('joining' particles) to the net growth rate ('joining' minus 'leaving'). Both the gross and net rates are calculated separately for each accretion mode, which are shown by the different curves in indicated in the legend. The most striking result here is that both the gross mass gain and the gross mass loss are comparable to, or even larger than, the {\it net} halo growth rate.}
\vspace{0.5cm}
\label{f:joining_and_leaving}
\end{figure}

We can learn more about this 'cycle' of particles through different halos from Figure \ref{f:joining_and_leaving}. There it is shown that the rate at which particles join and leave their halo is similar to, or even higher than, the {\it net} growth rate, for each of the different modes. As a quantitative example, shown in Figure \ref{f:joining_and_leaving}, the rate of 'smooth' particles joining their halos at $z\approx0$ is $\approx4$ times higher than the net growth rate due to smooth accretion, i.e.~only $\approx33\%$ higher than the 'leaving' rate of particles that previously arrived smoothly. These numbers drop toward higher redshift, where the 'cycle' is less significant, e.g.~at $z\approx2$ the values are lower roughly by a factor of $2$ compared to those shown in Figure \ref{f:joining_and_leaving}. Note that all 'leaving' particles leave their halos smoothly, i.e.~not as part of a bound subhalo, as such events have already been cleaned at the time of merger tree construction. Thus the red curve in Figure \ref{f:joining_and_leaving} shows that there are many particles that arrive via mergers and then stripped off of their subhalos inside the main halos and later leave the main halo 'smoothly'. It seems reasonable to suggest that this cycle is driven, at least partly, by fluctuations of particles that reside close to the halo boundary in and out of the region defined as the halo by the FOF algorithm. This is the reason we focus throughout the paper on the net growth rate. We leave a more detailed study of this cycle to future work.

\subsection{Alternative trees and halo definitions}
\label{s:alternative_definitions}
\begin{figure}[tbp]
\centering
\includegraphics[]{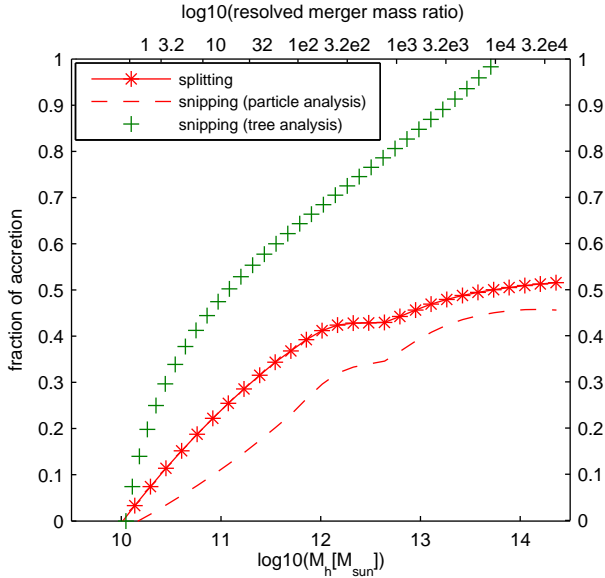}
\caption{Comparison of the contribution of the different accretion modes at $z=0.3$ in the "splitting" and "snipping" trees from the USM simulation. In the "snipping" trees ({\it dashed}) the contribution of mergers is {\it not} higher than in the "splitting" trees ({\it solid, asterisks}), unlike the results of the merger tree analysis ({\it pluses}, same as in Figure \ref{f:growth}). This demonstrates that merger trees with fragmentation events are not appropriate for studying halo growth modes, as they are not self-consistent. In fact, in the particle analysis the "snipping" merger contribution is somewhat lower than the "splitting" merger contribution due to an artificial effect caused by the snipping of the main progenitor trunk, see details in the text. For the "splitting" trees, there is very good agreement between the USM simulation ({\it solid, asterisks}) and the milli-Millennium simulation (Figure \ref{f:particle_net_accretion}).}
\vspace{0.5cm}
\label{f:particles_split_vs_snip}
\end{figure}

In Figure \ref{f:particles_split_vs_snip} we compare the merger contribution to the accretion rate from our "splitting" and "snipping" \citep{FakhouriO_07a} trees. The "snipping" algorithm is the most extreme case of leaving all fragmentations in the trees so that in the merger tree analysis the merger contribution does not converge ({\it pluses}, same as in Figure \ref{f:growth}). In contrast, the "snipping" merger contribution does not exceed $60\%$ when the {\it particle histories} are used ({\it dashed}). In fact, it is somewhat lower than in the "splitting" case ({\it solid, asterisks}) because, as we describe in \S\ref{s:particles_method}, the role of 'stripped' particles is overestimated at the expense of the other accretion modes. This is because the particles of 'snipped' halos, which lose their main progenitor trunks, are tagged 'stripped' instead of their original accretion modes. To verify this we divide the 'stripped' particles into two categories: 'just stripped' and 'stripped in the past'. The former category includes particles that at $s_{acc}-1$ belong to a halo that is not a progenitor of $h_{acc}$, and the latter category includes particles that were found to belong to a halo at an earlier time in the past. We find that 'just stripped' particles are negligible in the "splitting" trees but they become very significant, especially for low mass halos, in the "snipping" trees, because the particles of a 'snipped' halo $h_{acc}$ are identified at $s_{acc}-1$ in the artificially-connected FOF group, which is not a progenitor of $h_{acc}$. Hence, we see that the results of the analysis we present in this Section are not entirely merger tree-independent. The results for trees with remaining fragmentations are different depending on the analysis method (merger tree only or particle history analysis) and both suffer from artificial effects because the trees are not self-consistent. On the other hand, the analyses based on the "splitting" trees are consistent with each other (as already shown in Figure \ref{f:particle_net_accretion}).

\begin{figure}[tbp]
\centering
\includegraphics[]{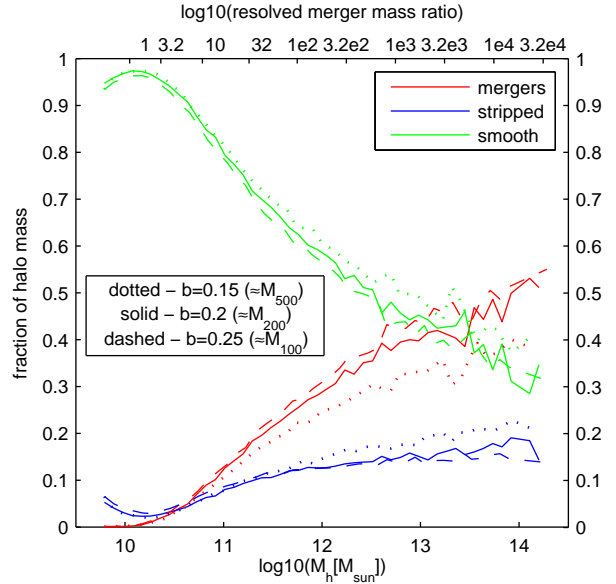}
\caption{The contribution of the different halo growth modes to the $z=0.3$ halo mass from the USM simulation. The results with the standard FOF group finder ({\it solid}) agree very well with the results from the milli-Millennium simulation (Figure \ref{f:halo_content}). The results based on different halo definitions (FOF groups with $b=0.25$, {\it dashed} and with $b=0.15$, {\it dotted}) are also virtually unchanged.}
\vspace{0.5cm}
\label{f:USM_M100_M200_M500}
\end{figure}

Finally we explore the sensitivity of our results to the halo definition by repeating our analysis with a different definition for a halo. The first step of structure identification, the FOF group finder, is run with different linking lengths of $b=0.25$ ($b=0.15$), corresponding approximately to structures with overdensities of $100$ ($500$) rather than $200$. This results in FOF groups that are on average $\approx15\%$ more ($\approx25\%$ less) massive. We then run SUBFIND and our splitting algorithm as before. The halos in our final catalogues are also more (less) massive compared with our original catalogues by similar factors. This is because SUBFIND is run on each FOF group separately and is restricted to working on particles within FOF groups only\footnote{In the $b=0.25$ case, where particles have been {\it added} to the FOF groups compared with the $b=0.2$ case, they are typically infalling onto their halos, i.e.~they have negative energies relative to the halo centers and are therefore considered bound by SUBFIND and included in the new halo catalogues.}. Figure \ref{f:USM_M100_M200_M500} shows the contribution of the different accretion modes to the halo mass at $z\approx0.3$ from the USM simulation for the three different FOF group definitions. The results agree very well with each other and with the results presented above from the Millennium Simulations. This agreement reflects the fact that the change of halo definition is roughly mass independent \citep{JenkinsA_01a,WhiteM_02a}, and so both the merger mass contribution and the total halo mass change in a similar way. It then follows that the smooth accretion mass contribution increases by the same factor and the relative contributions do not change. Note, however, that using a mass definition that changes the mass of halos as a function of mass (e.g.~\citealp{CuestaA_08a}) could lead to different results. For example, \citet{WarrenM_06a} suggested a correction to FOF group masses that reduces the mass of poorly resolved (low mass) FOF groups, an effect that would work in the direction of an even higher contribution from smooth accretion.

There exist a few alternatives to the FOF algorithm for identifying dark matter halos. In this paper we use only the FOF algorithm (with different linking lengths, as described above), but we believe that other halo definitions will not change the main results. One supporting evidence for that is the work of \citet{StewartK_08a}, who used the 'Bound Density Maxima' algorithm \citep{KlypinA_99a} for halo identification, and found results that agree well with ours. Another alternative would be to use 'Spherical Overdensity' halos, i.e.~defining the spherical region inside the virial radius as the halo. Unless FOF groups are organised in a way that the smoothly accreted mass is outside the virial radius and the merger accreted mass is inside, significant deviations from our results should not arise. The case is probably the opposite, because when FOF groups have very aspherical shapes it is mainly because a few substructures are connected together, i.e.~it is not that the outskirts of FOF groups constitute mainly of smoothly accreted material.

\section{Implications for galaxy formation}
\label{s:galaxy_formation}
Many theoretical models focus on the role of major mergers in galaxy evolution. In this paper we have shown that it is actually minor mergers ($x>10$) and smooth accretion that dominate halo growth by accounting for $\approx70\%$ of the accretion rate onto halos. Therefore, halos grow mainly continuously rather than in bursts with short duty cycles. This mode of accretion is more favourable for disk formation at all redshifts, and may in particular help in understanding galaxies at high redshift that show extreme star-formation rates but no signs of major mergers (e.g.~\citealp{GenzelR_08a,GenelS_08a,FoersterSchreiberN_09a}).

If there is a truly smooth accretion component, simulations from cosmological initial conditions naturally include it. Indeed, our result is consistent with hydrodynamical simulations that show that most baryonic accretion onto galaxies does not arrive in the form of mergers \citep{MuraliC_02a,SemelinB_05a,MallerA_06a,DekelA_09a}. Our findings suggest that as the resolution of future simulations increases, the mass contribution of small halos to the formation of galaxy-size halos will hardly increase further. This implies that $\approx40\%$ is a strong lower-limit on the mean fraction of pristine IGM gas in the baryons accreted onto halos of any given mass or redshift, since this smooth gas was never bound to any subhalos and is therefore not expected to have formed stars or to have become significantly enriched with metals, regardless of the star-formation efficiency and history of the baryons in the merging halos. It may also be expected that this $\approx40\%$ did not experience feedback from star-formation in smaller halos, and is likely "cold" when finally accreted onto a halo, with $T\gtrsim10^4K$ set by photoheating of IGM gas.

In the context of the "cold flow" mode of gas accretion onto galaxies \citep{BirnboimY_03a,Dekel06_a,KeresD_05a}, our results suggest that roughly half of the incoming gas is not clumpy, thus processes like tidal stripping and dynamical friction are irrelevant for it\footnote{Although it is possible that the gas becomes clumpy inside the halo on its way to the galaxy due to different instabilities (e.g.~\citealp{FieldG_65a,BurkertA_00a,MallerA_04a,KeresD_09b,BirnboimY_09a}).}. Theoretical or semi-analytical galaxy formation models, for example EPS-based models, may benefit from taking the smooth accretion component into account, as many properties of galaxies, such as size and morphology, depend on the way they accrete their baryons.

\begin{figure}[tbp]
\centering
\subfigure[]{
          \label{f:fraction_below_UV_threshold_z1}
          \includegraphics[]{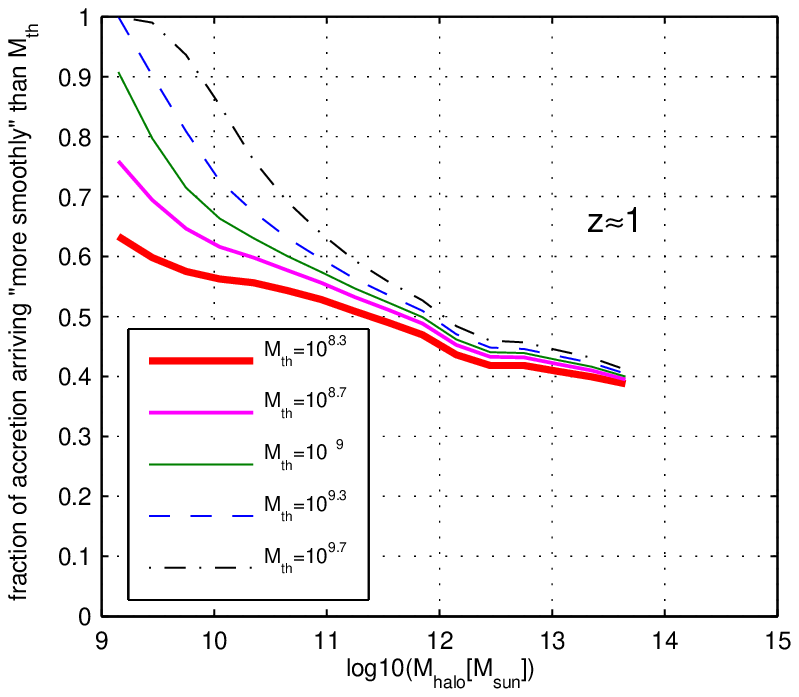}}
\subfigure[]{
          \label{f:fraction_below_UV_threshold_z0}
          \includegraphics[]{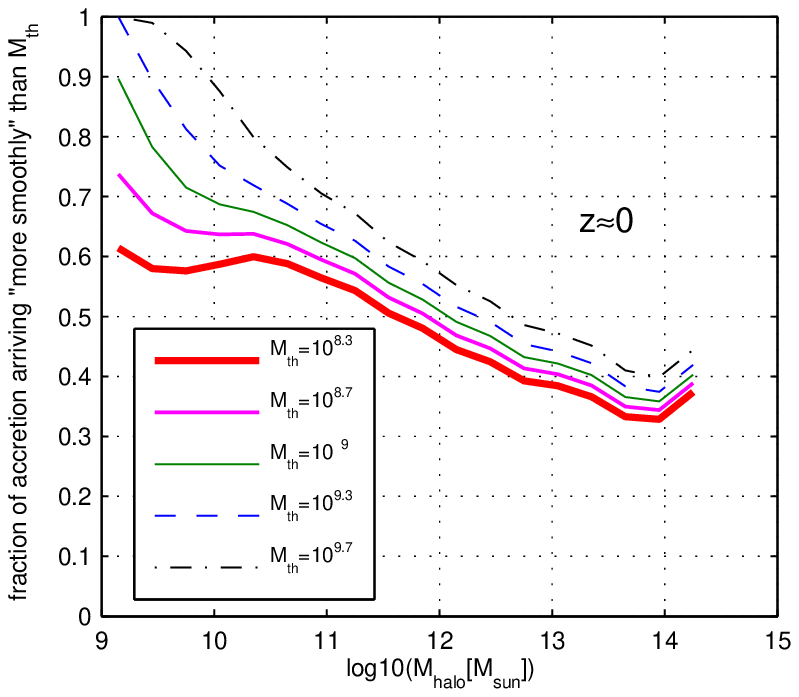}}
\caption{The fraction of mass accretion that arrives "more smoothly than" $M_{th}$, i.e.~by accretion of halos with $M<M_{th}$ plus smooth accretion. This fraction is shown as a function of halo mass, and for different values of $M_{th}$. These results are from the MS2 and are based directly on the merger trees, without using the fitting formula Equation \ref{e:rate_desc} or any extrapolation below the simulation's resolution limit. The results are plotted for accretion at $z\approx1$ ({\it top}) and $z\approx0$ ({\it bottom}). The resemblance of the $z\approx0$ results to those at $z\approx1$ highlights the fact that despite of an overestimation of the minor merger rate at $z\lesssim0.5$ in the "splitting" trees, the contribution of minor mergers to the total mass growth is still very small.}
\vspace{0.5cm}
\label{f:fraction_below_UV_threshold}
\end{figure}

Even if an extrapolation of the power-law index of the merger rate beyond $x=10^5$ is not valid and mergers of higher mass ratios do make up for the "missing" $\approx40\%$, our results are still significant. Since the MS2 resolves all halos with $T_{vir}>10^4K$ at $z\lesssim3$ ($M\gtrsim1.9\times10^{8}\Msun$), we do resolve most of the accretion of halos that have experienced star-formation, that is if star-formation is prohibited in halos with $T_{vir}<10^4K$ (e.g.~\citealp{ReesM_86a,EfstathiouG_92a,OkamotoT_09a,HoeftM_10a}). Current models of galaxy formation derive, assume or require strong suppression of cooling and/or star-formation below a virial temperature threshold that is even higher than $10^4K$ (e.g.~\citealp{BoucheN_10a,KravtsovA_10a}, and references therein). The fraction of the mass that arrives either smoothly or by accretion of small halos can be read off Figure \ref{f:growth} as the complement of the merger contribution. However, for the convenience of the reader we explicitly show in Figure \ref{f:fraction_below_UV_threshold} the fraction of accretion that arrives smoothly, for different values of a threshold mass below which accretion of {\it baryons} is assumed smooth. The results shown in Figure \ref{f:fraction_below_UV_threshold} do not depend on any extrapolation but are based directly on the MS2 (since all $M_{th}$ we use are resolved by the MS2). For example, by assuming that pre-heating from the cosmic UV background evaporates halos below $M_{th}\approx10^9\Msun$ ({\it green}), we infer that $10^{11}\Msun$ halos at $z\approx1$ ($z\approx0$) accrete $\approx55\%$ ($\approx65\%$) of their baryonic mass in smooth $T\gtrsim10^4K$ gas. Since $10^{11}\Msun$ halos do not form a stable virial shock and the cooling times are short, over-efficient galaxy formation in such halos is probably prevented by strong baryonic feedback processes.

\section{Comparison to previous work}
\label{s:previous_work}
Merger trees are regularly treated in the literature such that each node (whether a halo or a subhalo) never has more than one descendant. This constraint is motivated by the idea of hierarchical buildup of structure, but in practice, when standard halo definitions are used, fragmentation events are common too. To treat the formation of dark matter halos self-consistently, there are two alternatives. First is to allow for more than one decendant and consequently quantify both the merger rate and the fragmentation rate. Alternatively, the trees may be rearranged so that fragmentations do not exist. In this paper we use a variant of the second possibility, namely our "splitting" algorithm\footnote{In G09 we found that there is a good physical basis for our specific choice of a "splitting" algorithm, since fragmentations usually occur before the halos have had significant dynamical interaction.}. Our result agrees with that of \citet{StewartK_08a} and \citet{StewartK_08b}, as described in \S\ref{s:halo_growth}, since they use fragmentation-free merger trees, but disagrees with the results of \citet{FakhouriO_07a} and \citet{FakhouriO_08b}, who do not.

Recently \citet{FakhouriO_09a} found that the relative importance of mergers to the growth of halos in the MS correlates with large-scale environment. This, as they suggest, is further compelling evidence for the true diffuse nature of non-merger halo growth, even if the quantitative proportions they find from their trees differ from ours.

\citet{GuoQ_07a} investigated the relative growth of FOF groups in the MS via major mergers, minor mergers and 'diffused particles'. Their main findings that are different from our results are: (i) at large enough masses ($M\gtrsim3\times10^{12}\Msun$) merger accretion dominates over smooth accretion, (ii) the relative role of merger accretion, for a given mass, is larger at lower redshifts. Both differences occur because \citet{GuoQ_07a} consider "unprocessed" FOF groups, i.e.~implicitly use 'snipping' trees. They consider only instantaneous mass gain due to mergers but not instantaneous mass loss due to fragmentations, which becomes more important at low redshift and for high mass (better resolved) halos.

\citet{AnguloR_09a} studied halo formation histories in the framework of EPS assuming a finite dark matter particle mass. They found that if the dark matter particle is assumed to be a $100\GeV$ neutralino then Milky Way type halos are expected to have $\approx10\%$ smooth accretion, which is due to a minimum halo mass imposed by the free streaming of the dark matter particles. They compare their results to an N-body simulation, and find that up to a mass ratio $x=500$ the contribution of mergers is $\approx45\%$, in good agreement with our results. For the same mass ratio, they find the spherical collapse EPS model to give $\approx70\%$ accretion from mergers, in rough agreement with our results as well. The significant contribution they find for mergers with $x>500$ ($\approx25\%$ in the spherical collapse model) is due to the higher power-law index of the merger rate in EPS, as shown here in Figure \ref{f:EPS_ratio}. If instead the power-law index appropriate for very large $x$ is the one we find from the resolved regime in the Millennium Simulations, the contribution of mergers to halo growth practically vanishes even before reaching the free streaming mass.

\citet{MadauP_08a} report that less than $3\%$ of the final mass of the Via Lactea halo, which is a Milky Way-type halo simulated at high resolution, was accreted smoothly. This appears to stand in sharp contrast with our results, and there are a few possible explanations for this disagreement.
\begin{itemize}
\item
We do find Milky Way-type halos that have $<3\%$ of smooth accretion, but they are only approximately one in a thousand halos. It could then be that the Via Lactea halo is an untypical halo ($\approx3\sigma$) in this respect. This is a possible but undesired 'last resort' explanation, which we do not need to invoke, given the more likely explanation below.
\item
It could be that the higher resolution available for the Via Lactea halo is responsible for the difference. We suggest that this is {\it not} the dominant reason for the difference based on the following argument. It can be read from Figure $4$ in \citet{MadauP_08a} that halos of peak velocity ratio $<45$ have contributed to the Via Lactea halo $97\%$ of its final mass, which is all of the accretion that is associated with mergers. Such velocity ratios correspond approximately to mass ratios $x<45^3\approx10^5$, which can also be resolved for the most massive halos in the MS2. In other words, it seems that these are not the higher resolution mergers in the Via Lactea halo that contribute the "additional" (when compared with our results) $\approx37\%$ mass, but rather mergers that we {\it are} able to resolve. Moreover, we can consider just the {\it total} fraction reported by \citet{MadauP_08a}. The maximal mass ratio resolved by the Via Lactea simulation is $x\approx5\times10^6$, and our best resolved mass ratio is $x\approx10^5$. For this range $10^5<x<5\times10^6$ to account for $\approx97\%-60\%=37\%$ of the mass accretion, the power-law index of the merger rate would have to be $b\approx1.5$ at $10^5<x<5\times10^6$. Compared with the value $b\approx-0.3$ we find at $x<10^5$, this is an unlikely abrupt change.
\item
We believe the most probable explanation is that differences in analysis methods cause the results to be so different. \citet{MadauP_08a} sum up all the {\it peak} masses (over their full formation histories) of all halos that have merged into the main progenitor trunk and compare it to the mass inside $r_{200}$ at $z=0$. They do not account for the fact that some fraction of the mass of the merged subhalos ends up outside of $r_{200}$ at $z=0$, and so they overestimate the mass contribution of mergers.
\end{itemize}
To understand this issue with greater certainty, other high resolution simulations of individual halos should be further examined. To this end, J.~Wang et al.~2010, in preparation, analyse the high resolution Aquarius halos \citep{SpringelV_08a} with a method close to the one we use in \S\ref{s:particles} and find results that are more consistent with ours than with those of \citet{MadauP_08a}.

\section{Summary}
\label{s:summary}
In this paper we calculate the merger rates of dark matter halos and we investigate the role of smooth accretion versus mergers  to their growth. We extract the merger rates and accretion histories from the Millennium and Millenium-II simulations, combined with two additional, smaller, cosmological simulations. We use the "splitting" merger tree construction algorithm described in \S\ref{s:merger_trees_method} and in G09, and verify its reliability by reproducing our results by following individual particle histories alone, independent of merger tree construction algorithms.

We find that the contributions of all resolved mergers (up to mass ratios $\approx10^{5}$) to the total growth rate of halos do not exceed $60\%$, regardless of halo mass and redshift. Most of the {\it merger} contribution comes from small mass ratio ("major") mergers (e.g.~$1<x<10$ contribute $\approx30\%$ of the total growth), while "very minor" mergers add very little mass (e.g.~$10^{3}<x<10^{5}$ contribute just a few percent to the total halo growth). We find that the power-law index of the merger rate is such that if the merger rate is extrapolated beyond the maximum resolved mass ratio $\approx10^{5}$, the total contribution of {\it all} mergers saturates at $\approx60\%$. This suggests that a significant mass fraction of halos may be accreted in a genuinely smooth way.

Our results have important implications for galaxy formation models and the modes of baryonic accretion. If $\approx40\%$ of the dark matter that is accreted onto a halo was never previously bound in any merging smaller halos, then at least $\approx40\%$ of the baryons must also be accreted smoothly - as gas that was never previously heated by feedback processes or converted to stars. For the baryons, this $\approx40\%$ is a strong lower limit since halos with $T_{vir}<10^4K$ likely cannot retain their gas. The common assumption that halos with $T_{vir}<10^4K$ cannot retain their gas also makes our results insensitive to the limited resolution of the simulations we use, because we resolve all halos above this limit (at $z\lesssim3$). The implication is that a very large fraction of the baryonic matter falling into a halo must be pristine "cold" IGM gas, with $T\gtrsim10^4K$ set by IGM photoheating. This gas is not expected to have formed stars or to have become significantly enriched with metals, no matter what the star-formation efficiency and history of the baryons in any of the merging halos.

\acknowledgements
We thank Raul Angulo, Mike Boylan-Kolchin, Jerry Ostriker, Jie Wang and Simon White for advice and comments on the manuscript. We are grateful to Eyal Neistein for enlightening discussions and for kindly providing his code for computing EPS merger rates. We acknowledge the anonymous referee for a report that guided us towards significant improvement of the paper. The Millennium and Millennium-II Simulations databases and the web application providing online access to them were constructed as part of the activities of the German Astrophysical Virtual Observatory. We thank Gerard Lemson for helping us taking the most advantage of those databases, and Michaela Hirschmann for invaluable assistance with the USM simulation. We thank the DFG for support via German-Israeli Project Cooperation grant STE1869/1-1.GE625/15-1. SG acknowledges the PhD fellowship of the International Max Planck Research School in Astrophysics.

\end{document}